\documentclass[sigconf,nonacm]{acmart}
\usepackage{graphicx}
\usepackage{caption}
\usepackage{subcaption}
\usepackage{hyperref}
\usepackage{amsmath}

\title{Privacy-Preserving Bathroom Monitoring for Elderly Emergencies Using PIR and LiDAR Sensors}

\author{Youssouf Sidibé}
\affiliation{
  \institution{University of Michigan}
  \city{Ann Arbor}
  \country{USA}
}
\email{sidibey@umich.edu}

\author{Julia Gersey}
\affiliation{
  \institution{University of Michigan}
  \city{Ann Arbor}
  \country{USA}
}
\email{gersey@umich.edu}

\begin{document}

\begin{abstract}
In-home elderly monitoring requires systems that can detect emergency events—such as falls or prolonged inactivity—while preserving privacy and requiring no user input. These systems must be embedded into the surrounding environment, capable of capturing activity, and responding promptly. This paper presents a low-cost, privacy-preserving solution using Passive Infrared (PIR) and Light Detection and Ranging (LiDAR) sensors to track entries, sitting, exits, and emergency scenarios within a home bathroom setting. We developed and evaluated a rule-based detection system through five real-world experiments simulating elderly behavior. Annotated time-series graphs demonstrate the system's ability to detect dangerous states, such as motionless collapses, while maintaining privacy through non-visual sensing.
\end{abstract}


\maketitle

\section{Introduction}
The elderly population is rapidly growing; by 2040, over 78 million Americans will be age 65 or older \cite{census}. Many prefer to age autonomously, remaining in their own homes rather than relocating to assisted living facilities. However, aging in place comes with significant safety concerns, especially for elderly individuals who live alone. Bathrooms, in particular, are high-risk environments due to hard surfaces, slippery floors, and the physical demands of bending, sitting, and standing.

While traditional emergency systems like Life Alert exist, they rely on users being conscious and able to actively summon help—an assumption that often fails during real-world emergencies. According to the CDC, over 36 million falls are reported among older adults each year, resulting in more than 32,000 deaths \cite{cdc2022falls}. There is a critical need for systems that can passively monitor for emergencies without compromising user privacy to better support safe, independent living for the elderly.

Moreover, the global eldercare industry is valued at over \$1 trillion, with billions allocated specifically to remote monitoring, fall detection, and in-home aging technologies \cite{globenewswire2024, bcc2024}. The growing elderly population in the United States, Japan, and Europe highlights an urgent demand for scalable, affordable, and passive monitoring solutions that do not depend on manual activation \cite{globenewswire2024, patientone}.

In response to these needs, we present a privacy-preserving bathroom monitoring system that detects both falls and prolonged periods of inactivity using PIR and LiDAR sensors.

\section{Background and Related Work}

This work designs and deploys a privacy-respecting, always-on monitoring system that relies solely on motion and distance sensing.

Recent smart home systems and assistive technologies increasingly employ multimodal sensing to infer human activity. In retail spaces, for example, PIR sensors, RFID tags, and computer vision are commonly used to track motion, detect shoplifting, or optimize store layouts. RFID-based systems such as Xiao et al.'s item-level detection demonstrate that low-cost sensing can reliably monitor object states and movements, supporting the case for simple, scalable sensing architectures in monitoring tasks \cite{xiao2015rfid}. In residential environments, however, privacy becomes paramount—particularly in sensitive spaces like bathrooms. Frameworks for aging-in-place technologies emphasize the importance of non-invasive sensors, remote alert systems, and adaptive designs that support daily independence while preserving user dignity \cite{sciencedirect2025, ijmr2022}.

Camera-based solutions, while offering comprehensive coverage, introduce significant privacy concerns, especially in private spaces. Alternative approaches, such as stranger detection through structural vibration analysis, enable passive occupant monitoring without direct identification or video feeds \cite{dong2022stranger}. Similarly, monitoring gait health through footstep-induced vibrations has been explored for both elderly individuals and children with muscular dystrophy, demonstrating the viability of vibration-based methods for health monitoring \cite{fagert2019gait, dong2020mdvibe}.

Passive infrared (PIR) sensors offer a low-power, simple solution for motion detection, but provide only binary outputs and lack spatial resolution. LiDAR sensors, by contrast, emit laser pulses to measure distance and capture richer spatial information. Alone, each modality has limitations; however, when combined, they enable more nuanced inferences about a person's presence and activity. Recent work like MiLTON demonstrates the versatility of vibrometry for non-invasive sensing applications, motivating the use of multimodal approaches in privacy-sensitive environments \cite{gadre2022milton}. Vibration-based sensing platforms such as VibSense further highlight the potential of non-visual modalities for passive human interaction detection \cite{liu2017vibsense}.

In healthcare and smart home monitoring, multimodal sensing has been shown to improve robustness and reliability. Systems that fuse modalities such as radar, Bluetooth, and infrared sensing demonstrate enhanced emergency detection capabilities \cite{healthmonitor2021}. Occupant localization through footstep-induced vibrations, as presented by Mirshekari et al., provides a passive and scalable alternative to wearable-based systems \cite{mirshekari2018occupant}. In complex environments like retail or healthcare, multimodal fusion frameworks such as FAIM, which integrates vision and weight sensing, overcome limitations inherent to single-modality systems \cite{falcao2020faim}. Similarly, SenseTribute applies multimodal sensing to occupant identification tasks, achieving improved robustness and reduced false positives in smart home monitoring \cite{han2017sensetribute, han2018smarthome}. 

LiDAR's reliability in robotics applications, including obstacle detection and 3D mapping, further supports its suitability for structured environment monitoring. Drawing on these prior findings, this work combines PIR and LiDAR sensing with rule-based decision logic to create an explainable, modifiable, and privacy-preserving emergency detection system specifically tailored for bathroom environments.

\section{System Design}
\subsection{Hardware Setup}
Our system consists of three key components:
\begin{itemize}
  \item \textbf{PIR Sensor (HC-SR501)}: Detects motion based on IR fluctuations.
  \item \textbf{LiDAR Sensor (TF-Luna)}: Measures distance from sensor to closest object (range: 0.2–8m).
  \item \textbf{ESP32-S3 Microcontroller}: Collects data and streams via UART.
\end{itemize}

Both sensors are mounted at approximately 1.5 meters high, facing the toilet from a side wall. The PIR detects motion during entry or movement, while the LiDAR tracks the user’s distance from the wall, capturing transitions such as sitting, standing, exiting, or collapsing.

\begin{figure}[h]
  \centering
  \includegraphics[width=0.85\linewidth]{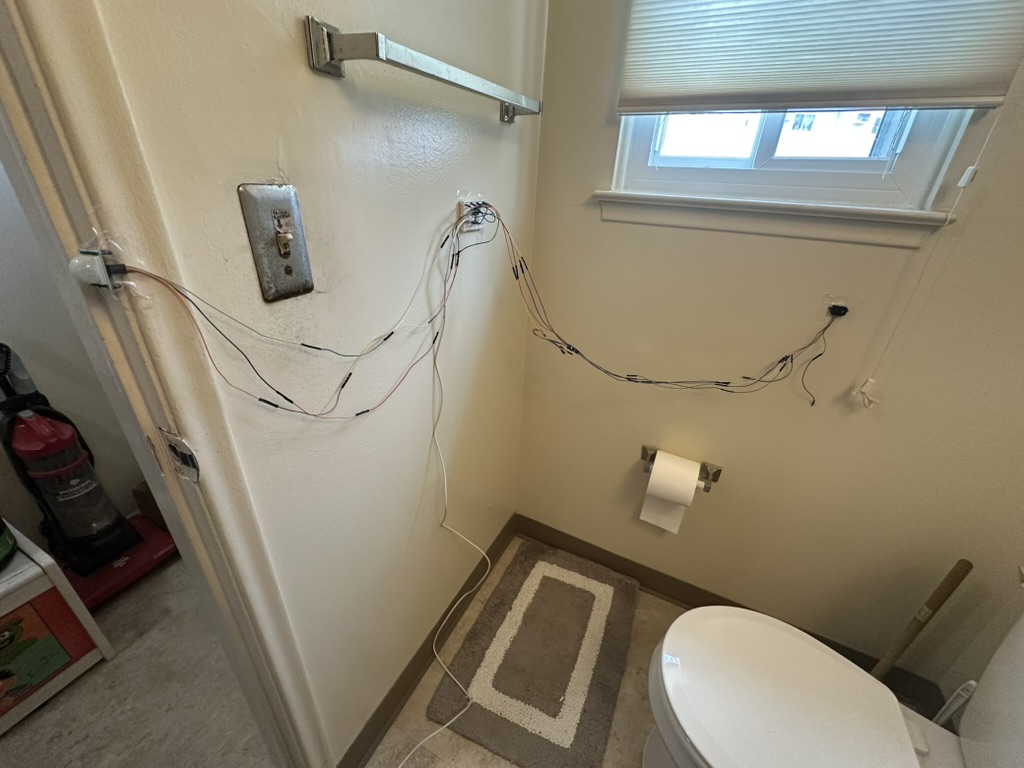}
  \caption{Experimental sensor setup in a test bathroom.}
\end{figure}

\subsection{State Logic}
The ESP32-S3 microcontroller polls sensor data every 50ms and logs it for analysis. Our rule-based logic maps sensor states to labeled activity phases:
\begin{itemize}
  \item \textbf{Entered}: PIR HIGH + LiDAR drop $<$ 150cm
  \item \textbf{Seated}: LiDAR stable between 30–80cm
  \item \textbf{Exited}: PIR HIGH + distance $>$ 130cm
  \item \textbf{Fall Suspected}: Distance $>$ 150cm \& no motion for 3+ minutes
  \item \textbf{Alert}: No motion \& distance $>$ 150cm for over 10 minutes
\end{itemize}

\section{Experiments and Results}
We ran five experiments simulating realistic elderly behavior. Data was logged via CoolTerm and plotted in Python.

\subsection{Experiment 1 – Normal Entry + Toilet Sit}
\textbf{Goal}: Test basic functionality. \textbf{Behavior}: Enter, sit for 1–2 minutes, exit. \textbf{Result}: Clean transition from “Entered” to “Seated” to “Exited” detected.

We placed the user in front of the LiDAR to observe entry detection. Once seated, distance readings stabilized between 60–70cm. Upon exiting, a sharp jump to 160cm and a final PIR HIGH confirmed the exit state. This scenario validates the system's ability to accurately track transitions between key behavioral states, which serves as a foundation for more complex emergency detection.

\begin{figure}[h]
  \centering
  \includegraphics[width=\linewidth]{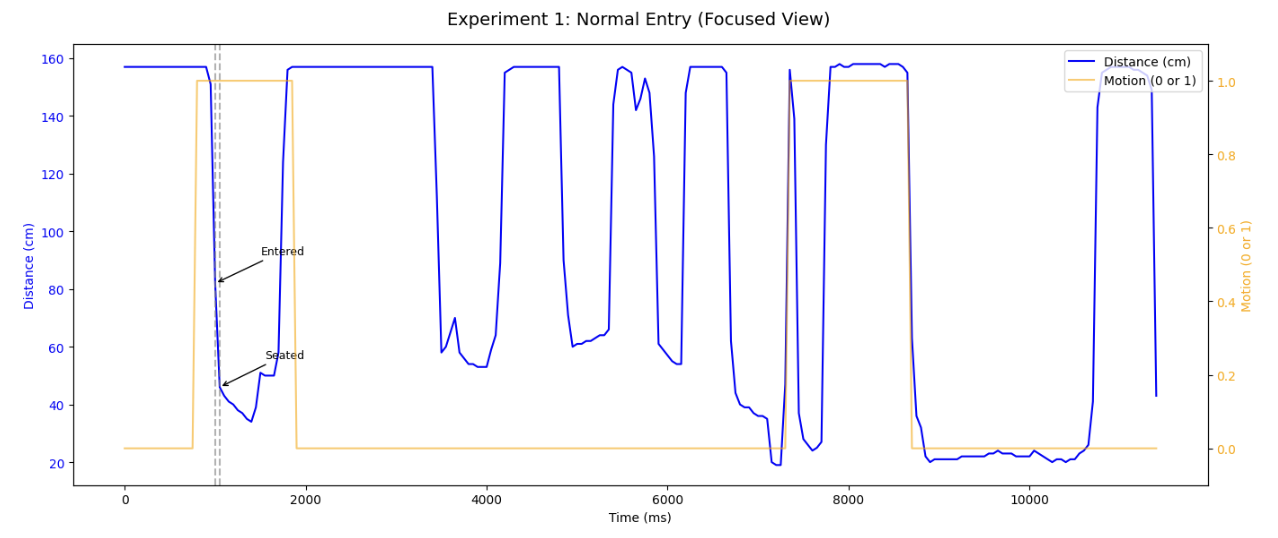}
  \caption{Experiment 1: Entry, sit, exit. Correct transitions detected.}
\end{figure}

\subsection{Experiment 2 – Long Sitting (No Motion 10+ min)}
\textbf{Goal}: Simulate risk of stroke, etc. \textbf{Behavior}: Remain seated still with no motion for >10 minutes. \textbf{Result}: PIR stayed LOW, LiDAR remained stable. System triggered “Warning” at 5 minutes, escalating to “Alert” at 10.

This scenario reflects a potential medical emergency such as fainting or a stroke while on the toilet. The system’s ability to escalate from a warning to an alert shows that it can differentiate passive presence from potential danger. No additional gestures or inputs were required from the user, demonstrating the advantage of fully passive sensing.

\begin{figure}[h]
  \centering
  \includegraphics[width=\linewidth]{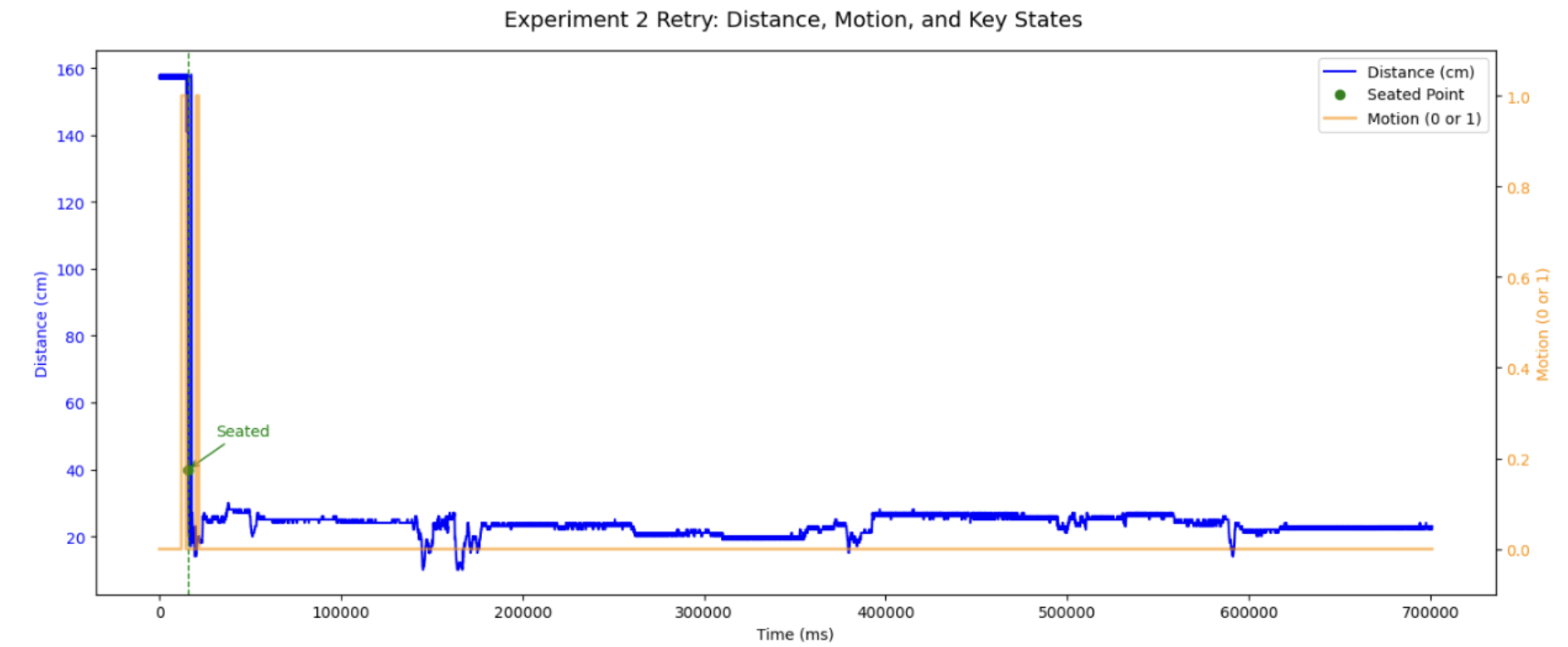}
  \caption{Experiment 2: Alert triggered after prolonged seated inactivity.}
\end{figure}

\subsection{Experiment 3 – Entry + Immediate Exit}
\textbf{Goal}: Avoid false alerts from quick visits. \textbf{Behavior}: Enter briefly, then leave. \textbf{Result}: PIR HIGH and LiDAR blip, but no state was sustained. \textbf{Interpretation}: System correctly ignores short visits.

This experiment is crucial to prevent alarm fatigue. We quickly entered the bathroom, triggering PIR and LiDAR drops, but exited within 20 seconds. Since no seated or motionless state was detected, the system returned to idle state as intended.

\begin{figure}[h]
  \centering
  \includegraphics[width=\linewidth]{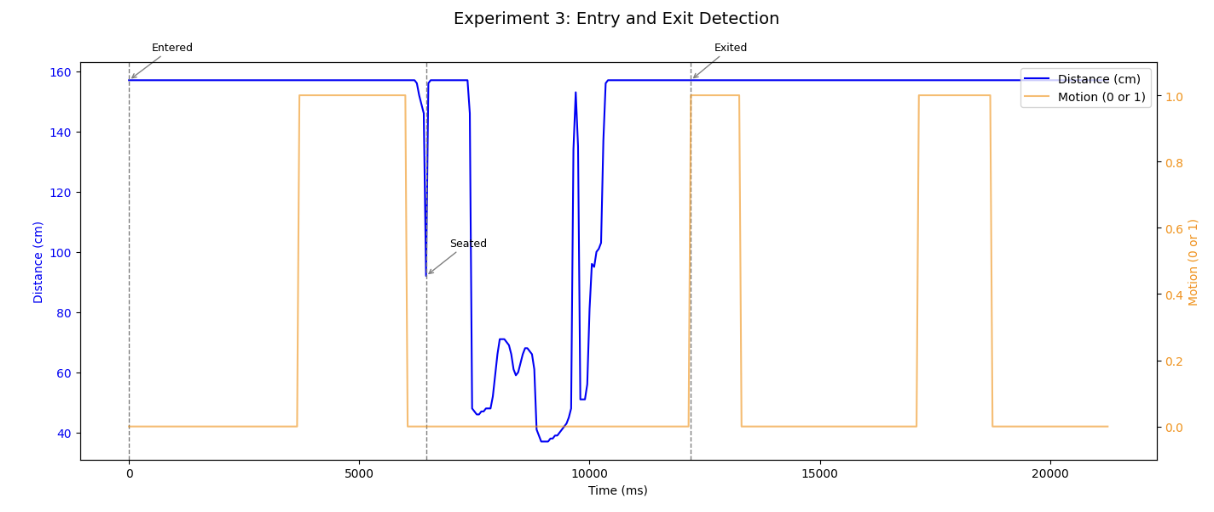}
  \caption{Experiment 3: No alert triggered. System correctly classifies short visit.}
\end{figure}

\subsection{Experiment 4 – Simulated Fall (Out of FOV)}
\textbf{Goal}: Detect fall beyond toilet. \textbf{Behavior}: Sit, then fall sideways. \textbf{Result}: LiDAR jumped, PIR LOW. “Fall Suspected” $\rightarrow$ “Alert.”

The user sat down and then slid out of the beam’s path. LiDAR spiked to 160cm, PIR stopped detecting motion. The system suspected a fall at 3 mins, then escalated to an alert.

\begin{figure}[h]
  \centering
  \includegraphics[width=\linewidth]{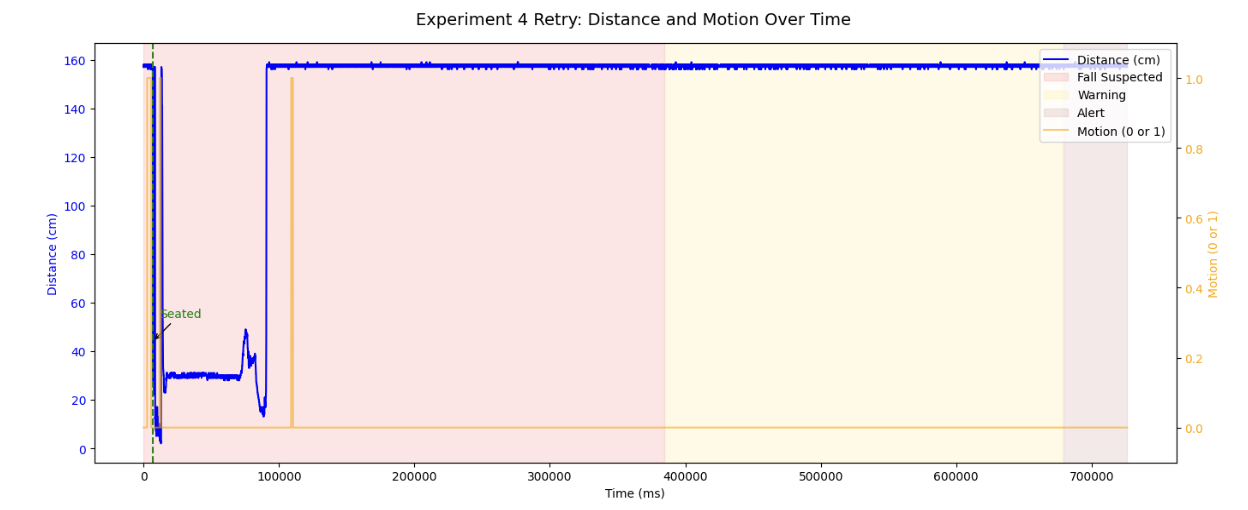}
  \caption{Experiment 4: Fall suspected after distance spikes and no motion.}
\end{figure}

\subsection{Experiment 5 – Collapse While Entering}
\textbf{Goal}: Detect early-stage fall before sitting. \textbf{Behavior}: Enter, then collapse. \textbf{Result}: PIR briefly HIGH, then LOW. Distance spikes. Alert at 10 min.

This tests one of the most realistic dangers—falling before even sitting. The PIR detects entry but goes LOW quickly. LiDAR briefly drops, then rises and remains static. The 10-minute timer leads to an alert trigger.

\begin{figure}[h]
  \centering
  \includegraphics[width=\linewidth]{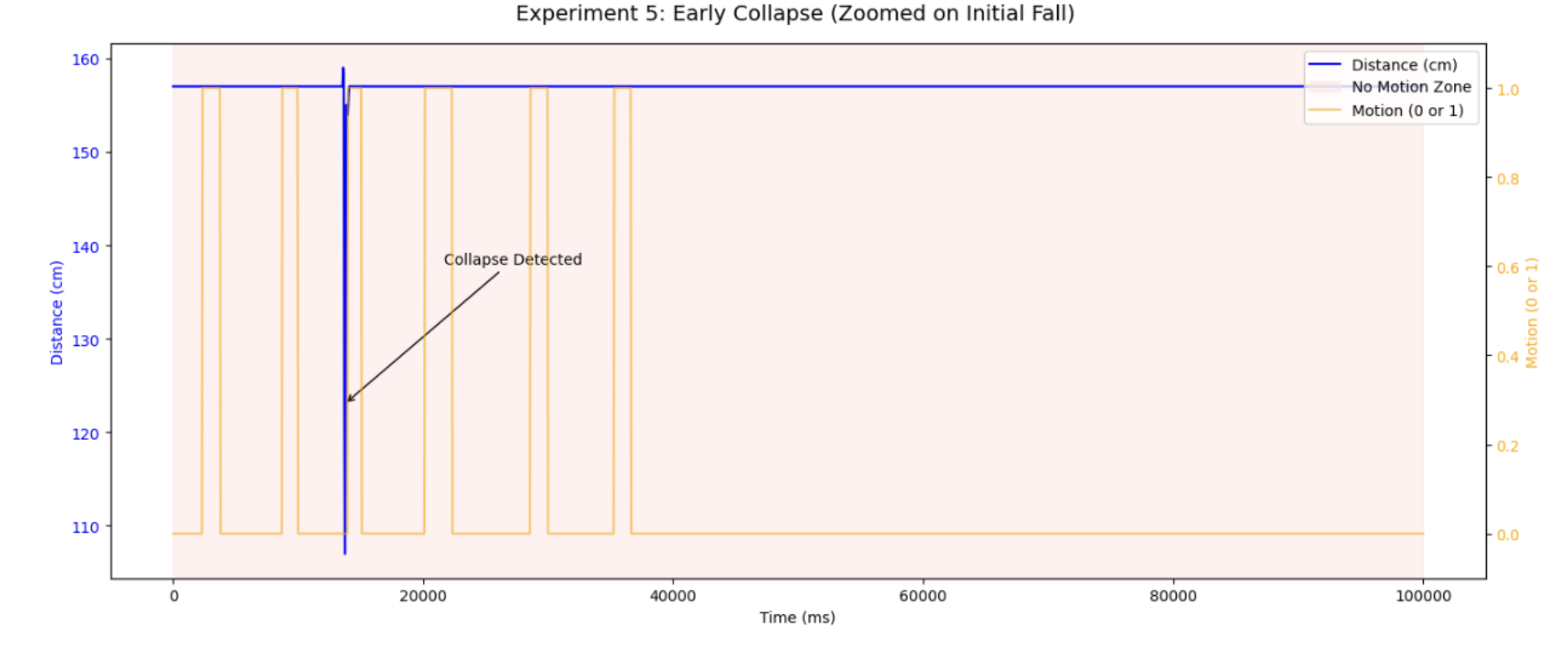}
  \caption{Experiment 5: Alert triggered after early collapse.}
\end{figure}

\section{Discussion}
Our multimodal setup balances simplicity, cost, and privacy. The experiments validate a system that is both robust and explainable. The threshold-based logic was sufficient to differentiate common cases, including subtle or dangerous ones.

\subsection{False Positives and Negatives}
False alerts can occur from long showers or meditation. Minor movement might delay alerts. Solutions include third sensor redundancy or modeling behavior patterns temporally. However, our tests showed that combining both PIR and LiDAR already reduces spurious signals significantly compared to using either modality alone.

\subsection{Why This System Matters}
Life Alert and similar solutions assume a user will press a button during distress. But many elderly users suffer cognitive decline or physical trauma that makes activation impossible. According to PatientOne, systems that operate independently of wearable devices are more likely to be adopted by elderly users with mobility issues or cognitive decline \cite{patientone}. Our system removes the responsibility of having to manage a wearable or manual device. It allows an elderly person to live independently without sacrificing safety. Complementary work like GaitVibe+ explores using floor vibrations for occupant localization and gait analysis, offering a non-visual method for monitoring movement patterns in-home \cite{dong2022gaitvibe}. Moreover, it's built on \$20 worth of sensors, with logic simple enough to run on a microcontroller—making it scalable in real-world homes. This direction aligns with a larger trend in aging-in-place innovation, which favors non-wearable systems for elderly populations with mobility or cognitive limitations \cite{metatech2024}. This design choice aligns with growing industry consensus that passive, non-contact sensing is more acceptable for in-home use than intrusive visual monitoring technologies \cite{sciencedirect2025, metatech2024}.

\subsection{Future Improvements}
BLE or Wi-Fi alerting, integration with smart rings or medical history logs, and ML-based classification are logical next steps. Sensor placement in other in-home locations (bedroom, hallway) could extend its use case.

\subsection{Deployment and Real-World Impact}
This system is designed to be both affordable and highly deployable. According to the Social Security Administration, over 12 million elderly Americans rely on these payments for more than 90\% of their income, making affordability a key barrier to entry for premium fall detection or medical response products. The total cost of the hardware stack—comprised of the ESP32-S3 microcontroller, a TF-Luna LiDAR module, and a PIR sensor—is under \$20.

Unlike subscription-based emergency services such as Life Alert, which can cost over \$50/month, this solution has no ongoing fees and does not rely on user compliance (e.g., wearing a pendant or bracelet). Elderly individuals often forget to wear such devices, or they may become physically unable to activate them in the case of a fall. Our sensor-based system removes that responsibility, instead shifting monitoring to the environment in a passive and automatic way. This promotes independence without sacrificing safety.

\subsection{Scalability and Market Context}
The aging-in-place technology market is expanding rapidly, with the global eldercare monitoring industry valued at over \$1 trillion \cite{credence, bcc2024, metatech2024}. Credence Research highlights the growth of non-contact fall detection systems, while BCC Research and MetaTech Insights report that caregivers increasingly prefer easy-to-install, passive sensor systems over camera-based solutions \cite{credence, bcc2024, metatech2024}. In countries like the US, Japan, and across Europe, populations over age 65 are growing faster than any other demographic. Meanwhile, the global eldercare and assistive device market continues to grow rapidly, supported by smart tech and monitoring innovation \cite{bcc2024}. As noted in Credence Research and MetaTech analyses, contactless, passive systems like mines align well with current market preferences for non-intrusive, easy-to-install solutions \cite{credence, metatech2024}. As demand for unobtrusive monitoring increases, simple solutions like mines provide a promising alternative to high-cost, privacy-invasive platforms. Bathroom-related falls account for a significant portion of emergency room visits among older adults. Early studies suggest that over 80\% of household falls among seniors occur in the bathroom—underscoring the urgency of monitoring such private, high-risk zones without relying on video.

Our implementation could be extended with BLE or Wi-Fi for caregiver alerts, or integrated with smart home hubs to trigger preprogrammed routines (e.g., unlocking the door for EMS, turning on lights, or sending location-aware texts to loved ones). The threshold logic used in our model can be trained with more complex behavior sequences, unlocking future applications such as activity recognition, daily routine tracking, or early illness detection.

\subsection{Policy Implications and Health Equity}
Beyond engineering impact, this project has the potential to reduce long-term healthcare costs by catching emergencies earlier. A fall left undetected for hours can lead to severe dehydration, pressure injuries, or worse. Early detection leads to better outcomes and less strain on healthcare systems. Future versions of this system could integrate with electronic health records (EHR), Medicare data, or insurance billing APIs to offer subsidized care pathways for low-income seniors. Policymakers and insurance providers are increasingly prioritizing aging-in-place technologies that reduce long-term care costs while remaining non-invasive and compliant with health data privacy laws \cite{globenewswire2024, bcc2024}.

This kind of passive monitoring system also aligns with HIPAA-compliant and GDPR-aligned health data practices. No visual or biometric information is recorded, reducing risks of surveillance misuse or legal liability for caregivers and institutions. It is explainable, transparent, and culturally considerate—making it better suited for large-scale deployment in both public and private eldercare contexts.

\section{Conclusion}
We developed and validated a privacy-preserving detection system for fall and inactivity monitoring in bathrooms by using PIR and LiDAR sensor. This low-cost, non-visual approach fills a critical gap in at-home eldercare with minimal user burden. It offers real-time insight while preserving the dignity and autonomy of elderly users.

This work advances the field of passive sensing and proposes a practical, ethical alternative to camera-based systems. It has the potential to scale across smart home deployments and contribute to aging-in-place solutions worldwide.

\bibliographystyle{ACM-Reference-Format}
\bibliography{main}

\end{document}